\documentclass[12pt,a4paper,final,fleqn,onecolumn]{article}

\usepackage{amssymb}
\usepackage{amsmath}

\usepackage[dvips]{graphicx}

\begin{document}

\noindent 

\noindent 

\noindent 

\noindent 

\noindent 

\begin{center}

\noindent \textbf{Optical contrast of single- and multi-layer graphene deposited on a gold substrate.}
\\*
\noindent I. Wlasny${}^{*1}$, P. Dabrowski${}^{1}$, Z. Klusek${}^{1}$

\noindent ${}^{1 }$Division of Physics and Technology of Nanometer Structures, Department of Solid State Physics, University of Lodz, Pomorska 149/153, 90-236 Lodz, Poland

\noindent * Corresponding author. E-mail adress: i\_wlasny@wp.pl (I. Wlasny)

\end{center}

\noindent \textbf{Abstract}\\

\noindent We study the optical contrast for single and multilayer graphene deposited on a Au/SiO${}_{2}$/Si substrate. Our results prove that optical microscopy allows for easy and quick localization, identification and counting of graphene layers. Visibility depends on the thicknesses of the Au and SiO${}_{2}$ layers as well as the wavelength of the light used. We propose optimum parameters for the detection of graphene.

\noindent \textbf{PACS:} 78.67.-n, 78.67.Wj, 78.66.-w , 78.66.Sq

\noindent \textbf{Keywords:} graphene, optical contrast, thin films\eject \textbf{1. Introduction}\\

\noindent Graphene is an allotrope of carbon, formed by single one-atom-thick layer of atoms. Ever since its first observations in 2004 [1], due to its exceptional electronic properties, such as ballistic transport of electrons [2, 3] on submicron scale or anomalous integer quantum Hall effect, it has been attracting growing interest. While free-standing graphene is a semi-metal (semiconductor with zero energy gap) it is known that its interface with a metal can alter its properties [4, 5]. Particularly, it has been proven theoretically (density functional theory - DFT) that the unique conical dispersion relation around K/K' points in graphene is preserved on (111) surfaces of Al, Cu, Ag, Pt, and Au [5]. However, this is accompanied with the change of position of the Dirac point (E${}_{D}$) relative to the Fermi level (E${}_{F}$) due to the presence of substrate (doping effect). This is especially important in the case of Au which is widely used in fabrication of metal-graphene contacts in graphene devices. The theoretical calculations have been recently confirmed by scanning tunneling microscopy/spectroscopy (STM/STS) results collected on mono-, bi- and tri- graphene\textbf{ }layer (MG, BG, TG) deposited on metallic conductive Au/SiO${}_{2}$/Si substrate [4]. This type of substrate enables to create a setup suitable for the STM/STS experiments without micro fabrication processes and studies of local electron density of states on multilayer graphene systems. Because of low throughput of STM at atomic resolution other means of finding, identification and estimation of the number of graphene layers are needed.  Most popular and easiest method involves measuring optical contrast which can be explained by standard thin film optics [6, 7]. 

\noindent \textbf{\eject 2. Methods}\\

\noindent Our calculations are based on geometry shown on Fig. 1. \\

\noindent 
\begin{center}
\noindent \textbf{\includegraphics*[width=2.37in, height=1.87in, keepaspectratio=false]{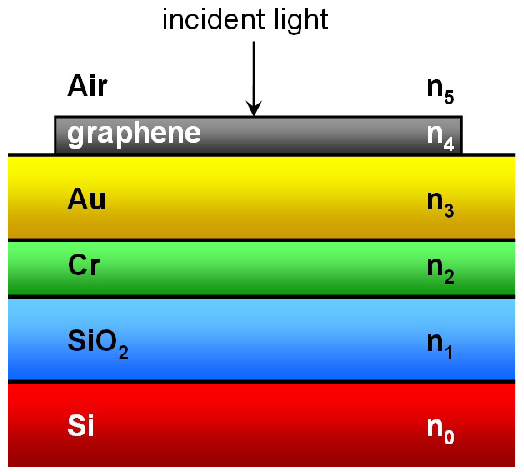}}

\noindent \textbf{Fig. 1} (color online) Arrangement of layers used in calculations.\\
\end{center}
\noindent 

\noindent We consider the case of light incident normally from air (\textbf{n${}_{5}$} = 1) onto four-layer structure consisting of graphene, Au, Cr, SiO${}_{2}$ and Si. As the index of refraction of graphene (\textbf{n${}_{4}$}~=~n${}_{1}$~--~\textit{i}~k${}_{1}$) we assume the index of refraction of graphite (2.6 -- 1.3 \textit{i}), which has been shown to be sufficient approximation [7]. Thickness of graphene is estimated as d${}_{4}$~=~[0.35~+~0.33~(t~-~1)] nm, where t is number of graphene layers [8]. Thickness of Au layer is given by d${}_{3}$, and as shown on Fig. 2a its refraction index (\textbf{n${}_{3}$}) is strongly dependant on wavelength of light [9]. 

\noindent As Au cannot be deposited directly on SiO${}_{2}$ an adhesive layer is needed. For this purpose Cr is commonly used. Thickness of adhesive layer is denoted as d${}_{3}$ and is fixed to 1 nm in all calculations. Refraction index of Cr (\textbf{n${}_{2}$}) is dependant on wavelength [9]. Its values are shown in Fig. 2b.

\noindent SiO${}_{2}$ layer is described by thickness d${}_{1}$. Its refraction index is also dependant on wavelength, but value \textbf{n${}_{1}$} = 1.47 accurately approximates it for visible light spectrum [7, 9]. Index of refraction of Si is assumed as \textbf{n${}_{0}$} = 5.6 -- 0.4 \textit{i} [7] and as in the case of SiO${}_{2}$ its dependence on wavelength can be omitted [7]. Si layer is assumed to be of semi-infinite thickness. 

\noindent In considered geometry, on each of the interfaces part of incoming wave is transmitted and the other part is reflected.\\
\begin{center}
\noindent \textbf{\includegraphics*[width=1.97in, height=2.91in, keepaspectratio=false]{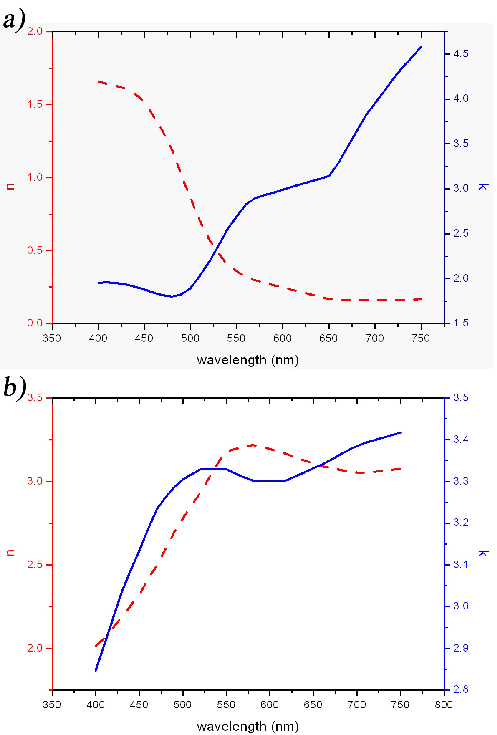}}

\noindent \textbf{Fig. 2} (color online) Real (n; red, dashed line) and imaginary (k; blue, solid line) parts of index of refraction of (a) Au (n${}_{3}$) and (b) Cr (n${}_{2}$) as a function of wavelength [9].\\
\end{center}
\noindent 

\noindent Interface between k and k+1 numbered layers is described by relative amplitude of reflected wave. This value is given by \eqref{GrindEQ__1_} [6]
\begin{equation} \label{GrindEQ__1_} 
r_{k} =\frac{n_{k+1} -n_{k} }{n_{k+1} +n_{k} }  
\end{equation} 
Optical path covered by light in k-numbered layer is defined as \eqref{GrindEQ__2_} [6]
\begin{equation} \label{GrindEQ__2_} 
\Delta _{k} =\frac{4\pi }{\lambda } n_{k} d_{k}  
\end{equation} 
Relative amplitude of reflected light from the system with k interfaces is described by equation \eqref{GrindEQ__3_}
\begin{equation} \label{GrindEQ__3_} 
re^{i\varepsilon } _{(k)} =\frac{r_{k-1} +re^{i\varepsilon } _{(k-1)} e^{-i\Delta _{k-1} } }{1+r_{k-1} re^{i\varepsilon } _{(k-1)} e^{-i\Delta _{k-1} } }  
\end{equation} 
Where e${}^{i}$${}^{\varepsilon }$ is phase and relative amplitude of system with two interfaces is given by \eqref{GrindEQ__4_} [6].
\begin{equation} \label{GrindEQ__4_} 
re^{i\varepsilon } _{(2)} =\frac{r_{1} +r_{0} e^{-i\Delta _{1} } }{1+r_{1} r_{0} e^{-i\Delta _{1} } }  
\end{equation} 
Reflectance of system with k interfaces is equal to \eqref{GrindEQ__5_} [6].
\begin{equation} \label{GrindEQ__5_} 
R_{(k)} \left(\lambda \right)\equiv \frac{I_{(k)} \left(\lambda \right)}{I_{0} \left(\lambda \right)} =\left|re^{i\varepsilon } _{(k)} \right|^{2}  
\end{equation} 
Where I${}_{0}$(l) is the intensity of incident light. Optical contrast of graphene for setup presented in Fig.1 is defined as \eqref{GrindEQ__6_}.
\begin{equation} \label{GrindEQ__6_} 
C\left(\lambda \right)=\frac{\left. I_{(5)} \left(\lambda \right)\right|_{n_{4} =1,d_{4} =0} -\left. I_{(5)} \left(\lambda \right)\right|_{n_{4} =2.6-1.3i,d_{4} \ne 0} }{\left. I_{(5)} \left(\lambda \right)\right|_{n_{4} =1,d_{4} =0} }  
\end{equation} \\
\textbf{3. Results and discussion}\\

\noindent Fig.3 shows plotted values of contrast of graphene as a function of SiO${}_{2}$ thickness and wavelength for selected thicknesses of Au layer. One can see that for thickness of Au layer lesser than 10 nm optical contrast can be observed for almost whole spectrum of visible light. Area of high contrast seems to disappear for high and low wavelengths with increasing Au layer thickness. Our calculations show that for d${}_{3}$ $>$ 30 nm no contrast can be observed for any thickness of SiO${}_{2}$ layer or wavelength. This is because light is only reflected at air/graphene and graphene/Au interfaces and light transmitted through the latter interface is absorbed almost completely. We believe that 5 -- 8 nm thick Au layer is most beneficial for purpose of graphene identification, as those values allow observation of high values of contrast (up to 60\%) while providing higher intensities of light due to less extinction. \\
\begin{center}
\noindent \includegraphics*[width=6.30in, height=2.93in, keepaspectratio=false]{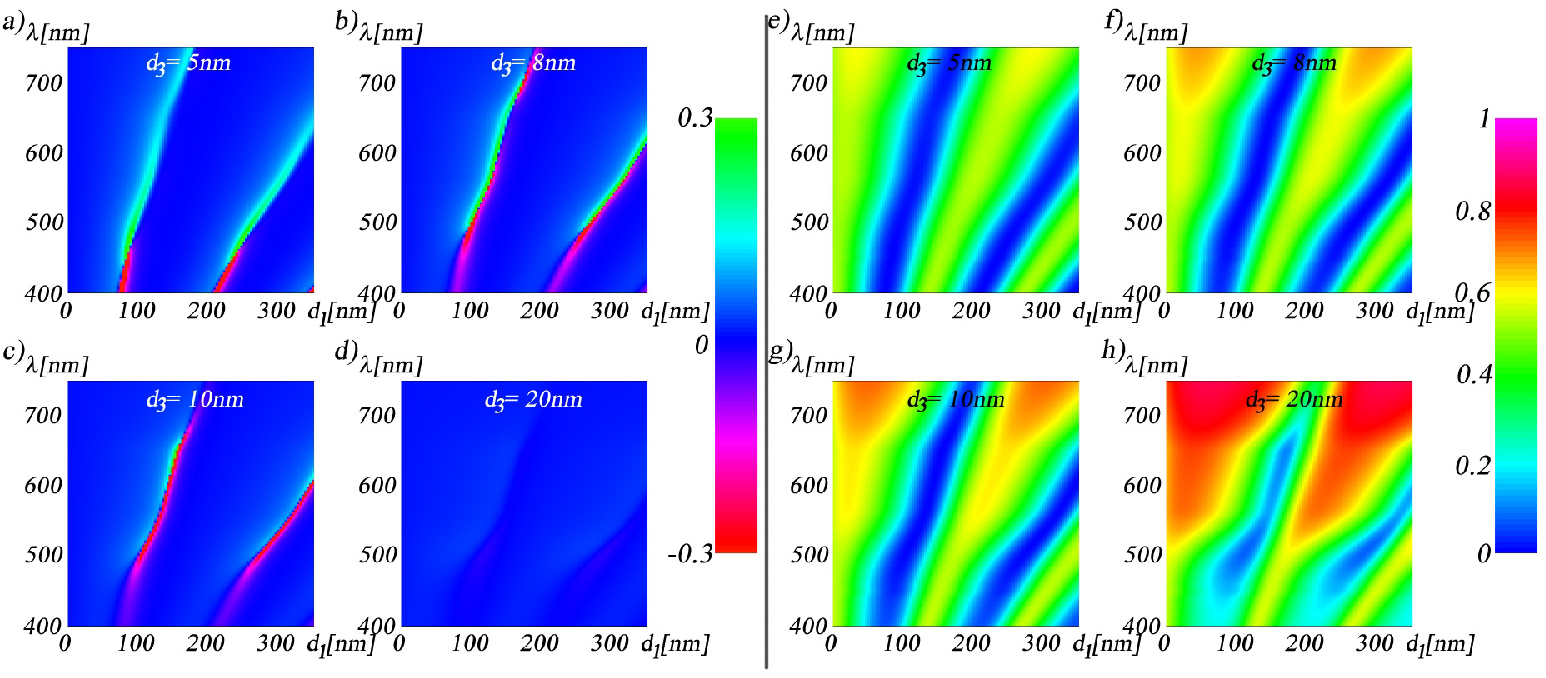}
\textbf{Fig. 3} (color online) Optical contrast (a -- d) and reflectance (e -- h) of graphene monolayer as a function of SiO${}_{2}$ thickness (in nm) and wavelength (in nm) for selected thicknesses of Au layer (a~--~5 nm, b -- 8 nm, c -- 10 nm, d -- 20 nm). Range of contrast values shown has been narrowed to enhance visibility of lesser values of contrast.\\
\end{center}
\noindent 

\noindent On Fig.3 a -- d one can notice that there are multiple areas of high contrast for any given wavelengths. However, first of these areas seem to best suited for observation of graphene, because optical contrast appears for wider range of wavelengths for fixed SiO${}_{2}$ thickness. While thickness of SiO${}_{2}$ layer ranging from 90 nm to 180 nm is suited for identification, green light is most comfortable for observer's eyes [7], thus range 90 nm -- 130 nm is desired.

\noindent While conducting observations of graphene on Au/Cr/SiO${}_{2}$/Si substrate one has to take into account dependency of intensity of reflected light on other parameters. Fig.3 e - h show that intensity of reflected light decreases with increasing value of contrast. Our calculations show that this effect appears for all values of d${}_{1}$ and d${}_{3}$. One has to remember to provide light source with high enough intensity. This dependency also has to be taken into account when light is provided through non-monochromatic filters.

\noindent Fig. 3 a - d Also show that at low wavelengths high values of negative contrast can be observed (graphene seem brighter than the substrate). In these areas contrast \eqref{GrindEQ__6_} is not well defined (the definition is kept for plot consistency). Values of well defined contrast in these areas are smaller than absolute values of negative, not well defined contrast.\\
\begin{center}
\noindent \includegraphics*[width=2.42in, height=1.41in, keepaspectratio=false]{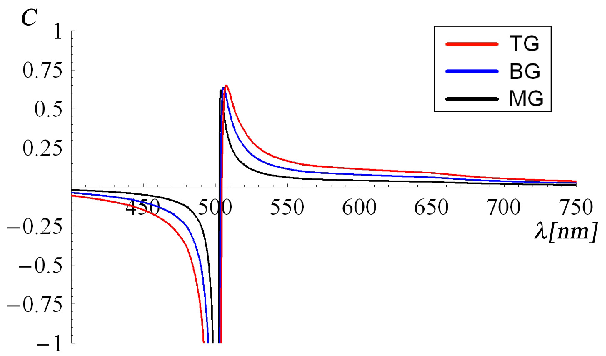}

\noindent \textbf{Fig. 4} (color online) Optical contrast of graphene as a function of SiO${}_{2}$ wavelength of light for d${}_{SiO2}$~=~100~nm and Au~=~8~nm for mono-, bi- and trilayer of graphene.
\end{center}
\noindent 

\noindent Fig.4 shows that mono-, bi- and trilayers of graphene can be easily distinguished between each other for nearly all wavelengths. Our results prove that optical contrast can be used to identify the number of graphene layers.

\noindent When comparing  our results with experimental data one has to remember that contrast can be different for angular incidence of light, as the optical paths inside each layers is longer [10, 11] and different polarizations have to be considered separately [6]. It is also important that parameters of deposition of each layer may have influence on their indices of refraction [12, 13]. \\

\noindent 

\noindent \textbf{4. Summary}\\

\noindent In summary by using extended Fresnel theory we have shown that optical contrast of graphene on Au/SiO${}_{2}$/Si substrate can be achieved for all wavelengths as well as almost all values of SiO${}_{2}$ thickness, but only for certain range of thickness of Au layer. We have found optimal values of thicknesses of each layer which are \~{}7 nm for Au and \~{}110 nm for SiO${}_{2}$. We have shown that such parameters are optimal for mono-, bi- and trilayer of graphene. \\

\noindent 

\noindent \textbf{Acknowledgments}

\noindent This work was ?nancially supported by Polish Ministry of Science and Higher Education in the frame of grant ``Investigations of electronic structure of graphene deposited on conductive and nonconductive surfaces by STM/STS/CITS/AFM and quantum electrodynamics N N202 204737.'' and from the European Social Fund implemented under the Human Capital Operational Programme (POKL), Project: D-RIM.

\noindent \textbf{}

\noindent \textbf{\eject References}\\

\noindent [1] K.S. Novoselov, A.K. Geim, S.V. Morozov, D. Jiang, Y. Zhang, S.V. Dubonos, Science 306, 669 (2004), DOI: 10.1126/science.1102896

\noindent [2] C. Neto, F. Guinea, N.M.R. Peres, K.S. Novoselov, A.K. Geim, Rev Mod Phys 81, 109 (2009), DOI: 10.1103/RevModPhys.81.109

\noindent [3] A.K. Geim, K.S. Novoselov, Nat Mater 6, 183 (2007), DOI: 10.1038/nmat1849 

\noindent [4] Z. Klusek, P. Dabrowski, P. Kowalczyk, W. Kozlowski, W. Olejniczak, P. Blake, et al., Appl Phys Lett 95, 113114 (2009), DOI: 10.1063/1.3231440

\noindent [5] G. Giovannetti, P. Khomyakov, G. Brocks, V.M. Karpan, J. van den Brink, P.J. Kelly, Phys Rev Lett 101, 026803 (2008), DOI: 10.1103/PhysRevLett.101.026803

\noindent [6] H. Anders, Thin films in optics. (Focal, CityplaceLondon, 1967)

\noindent [7] P. Blake, E.W. Hill, Appl Phys 91, 063124 (2007), DOI: 10.1063/1.2768624

\noindent [8] Z.H. Ni, H.M. Wang, J. Kasim, H.M. Fan, T. Yu, Y. Wu, et al., Nano Lett 7, 2758 (2007), DOI: 10.1021/nl071254m

\noindent [9] E.D. Palik. Handbook of Optical Constants of Solids. (Academic, placeStateNew York, 1991)

\noindent [10] V. Yu, M. Hilke, Appl Phys Lett 95, 151904 (2009), DOI:10.1063/1.3247967

\noindent [11] D.S.L. Abergel, A. Russell, V.I. Fal'ko, Appl Phys Lett 91, 063125 (2007), DOI: 10.1063/1.2768625

\noindent [12] Zhang G F, Zheng X, Guo L J, Liu Z T, Xiu N K. Influence of deposition parameters on the refractive index and growth rate of diamond-like carbon films. Surf Coat Tech 64, 127 (1994), DOI: 10.1016/S0257-8972(09)90013-5

\noindent [13] Borgogno J P, Flory F, Roche P, Schmitt B, Albrand G, Pelletier E, et al. Refractive index and inhomogeneity of thin films. Appl Opt 23, 2567 (1984), DOI: 10.1364/AO.23.003567

\noindent 

\noindent

\end{document}